\documentclass[reprint, showpacs,twocolumn,amsmath,amssymb,superscriptaddress, aps, prb]{revtex4}

\usepackage{graphicx}
\usepackage{dcolumn}
\usepackage{bm}
\begin{document}

\title{
Resistive detection of nuclear spins in a single quantum dot\\
under Kondo effect regime
}

\author{M. Kawamura}
	\affiliation{RIKEN Advanced Science Institute, Saitama 351-0198, Japan}
	\email{minoru@riken.jp}
	\affiliation{PRESTO, Japan Science and Technology Agency, Saitama 333-0012,Japan}
\author{D. Gottwald}
	\affiliation{RIKEN Advanced Science Institute, Saitama 351-0198, Japan}
\author{K. Ono}
	\affiliation{RIKEN Advanced Science Institute, Saitama 351-0198, Japan}
\author{T. Machida}
	\affiliation{Institute of Industrial Science,
		University of Tokyo, Tokyo 153-8505, Japan}
	\affiliation{Institute for Nano Quantum Information Electronics,
		University of Tokyo, Tokyo 153-8505, Japan}
\author{K. Kono}
	\affiliation{RIKEN Advanced Science Institute, Saitama 351-0198, Japan}
\date{\today}


\begin{abstract}
We study dynamic polarization
and resistive detection of nuclear spins
in a semiconductor quantum dot (QD) under the Kondo effect regime.
We find that the differential conductance spectra of the QD
exhibit hysteresis under the Kondo effect regime in magnetic fields.
Relevance of nuclear spins to the hysteresis 
is confirmed by the detection of 
nuclear magnetic resonance signals
by monitoring the differential conductance.
We attribute the origin of the hysteresis
to the dynamic nuclear spin polarization  (DNP) induced in the QD.
Using the DNP,
we demonstrate nuclear spin relaxation rate measurements
in the QD under the Kondo effect regime.
\end{abstract}

\maketitle

The Kondo effect,
arising from the interaction between a localized spin and itinerant spins at the Fermi energy,
is one of the most fundamental many-body phenomena
observed in semiconductor quantum dot (QD) systems\cite{Meir, Goldhaber-Gordon, Cronenwett, vanderWiel}.
The differential conductance (${\rm d}I/{\rm d}V_{\rm sd}$) spectrum of the QD exhibits 
a sharp zero-bias conductance peak (ZBCP),
which reflects the Kondo resonance at the Fermi energy.
In the presence of an external magnetic field  $B$,
the energy for the Kondo resonance shifts away from the Fermi energy by $\pm |{\rm g}^{*}|\mu_{\rm B}B$
due to the Zeeman splitting of the electronic state in the QD\cite{Meir},
where ${\rm g}^{*}$ is the effective g-factor for electrons and 
$\mu_{\rm B}$ is the Bohr magneton.
Consequently, the ${\rm d}I/{\rm d}V_{\rm sd}$ spectrum
splits into two peaks at finite bias voltages.
The peak-to-peak voltage $V_{\rm p-p}$, defined as the difference in the bias voltage
between the split differential conductance peaks, is given
by twice the Zeeman  energy\cite{Meir, Cronenwett}
 $V_{\rm p-p} = 2|{\rm g}^{*}|\mu_{\rm B}B/e$.

Because an electron spin $\vec{S}$ in a semiconductor device
can interact with a nuclear spin $\vec{I}$ of the host material through
the contact hyperfine interaction
$H_{\rm hyp}= A\vec{I}\cdot \vec{S} = A(I^{+}S^{-}+I^{-}S^{+})/2 + AI_{z}S_{z}$,
where $A$ is the hyperfine coupling constant,
a dynamic nuclear spin polarization (DNP) can be build
by driving electron spins using optical or electrical means\cite{Abragam, Slichter}.
The polarized nuclear spins modify the Zeeman  energy for electrons
through an effective magnetic field
 $B_{\rm N}  = A\langle I_{z}\rangle /{\rm g}^{*}\mu_{\rm B}$ [Fig.~1(a)].
In a QD under the Kondo effect regime,
the modification of the Zeeman energy is expected to
change the peak-to-peak voltage $V_{\rm p-p}$
between the split ${\rm d}I/{\rm d}V_{\rm sd}$ peaks
to $V_{\rm p-p} = 2|{\rm g}^{*}|\mu_{\rm B}(B + B_{\rm N})/e$,
as shown schematically in Fig.~1(b).
Therefore a small change in the nuclear spin polarization
can be sensitively detected by measuring the ${\rm d}I/{\rm d}V_{\rm sd}$ spectrum.
However, to the best of our knowledge, 
effects of  nuclear spins on transport properties of a QD
under the Kondo effect regime have not been addressed experimentally.
In addition, establishment of a novel technique for the resistive detection
of nuclear spins has a potential
to open a way for studying electron spin properties in a QD
through nuclear magnetic resonance (NMR) measurements.

\begin{figure}[t]
	\begin{center}
		\includegraphics[width=8.5cm]{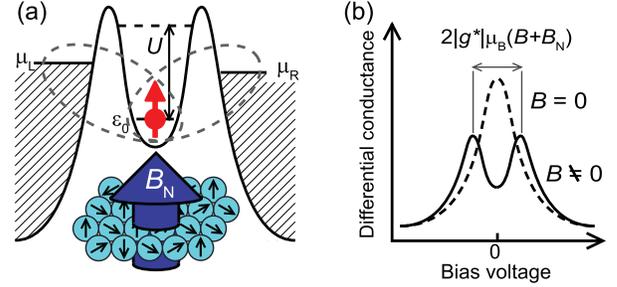}
		\caption{
			\label{fig1}
			(color online).
			(a)
			Schematic of a QD with nuclear spins.
			An electron spin in the QD is influenced by an effective magnetic field $B_{\rm N}$ produced by the nuclear spins.
			(b)
			Schematic of differential conductance spectra in a QD under the Kondo effect regime.
			Application of an external magnetic field $B$ causes
			the splitting of the ${\rm d}I/{\rm d}V_{\rm sd}$ spectrum.
			The peak-to-peak voltage between the split ${\rm d}I/{\rm d}V_{\rm sd}$ peaks
			is further modified by $B_{\rm N}$.
			}
	\end{center}
\end{figure}

In this paper,
we report dynamic polarization and resistive detection of nuclear spins in 
a single QD under the Kondo effect regime.
We find that the ${\rm d}I/{\rm d}V_{\rm sd}$ spectra
exhibit hysteresis under the Kondo effect regime in magnetic fields.
We also find that the value of ${\rm d}I/{\rm d}V_{\rm sd}$ evolves slowly
at the bias voltages between the split ${\rm d}I/{\rm d}V_{\rm sd}$ peaks.
Relevance of  nuclear spins to the hysteresis 
and the slow evolution of ${\rm d}I/{\rm d}V_{\rm sd}$
is confirmed by the detection of NMR signals by monitoring ${\rm d}I/{\rm d}V_{\rm sd}$.
We attribute the origin of the hysteresis and the slow evolution of ${\rm d}I/{\rm d}V_{\rm sd}$
to the DNP induced in the QD.
Using the DNP,
we demonstrate nuclear spin relaxation rate measurements
in the QD under the Kondo effect regime.

We prepared QDs using 
a wafer of GaAs/Al$_{0.3}$Ga$_{0.7}$As single heterostructure with 
a two-dimensional electron gas (2DEG) at the interface.
The carrier density and mobility of the 2DEG are
$n$ = 2.3 $\times$10$^{15}$~m$^{-2}$ and $\mu$ = 17~m$^2$/Vs, respectively.
The QDs were made using split-gate devices [inset of Fig. 2(e)],
following the approach in Refs.~\onlinecite{Sfigakis} and \onlinecite{Klochan};
QDs form occasionally in split-gate devices near the pinch-off conditions
in the presence  of disorder\cite{McEuen},
and some of them exhibit the Kondo effect\cite{Sfigakis, Klochan}.
The devices were cooled using a dilution refrigerator
and magnetic fields were applied parallel to the plane of the 2DEG.
A standard lock-in technique with 
an excitation voltage of 10~$\mu$V and a frequency of 18~Hz
was used to measure ${\rm d}I/{\rm d}V_{\rm sd}$ applying 
a bias voltage $V_{\rm sd}$ between the source and drain electrodes.

\begin{figure}[t]
	\begin{center}
	\includegraphics[width=8.5cm]{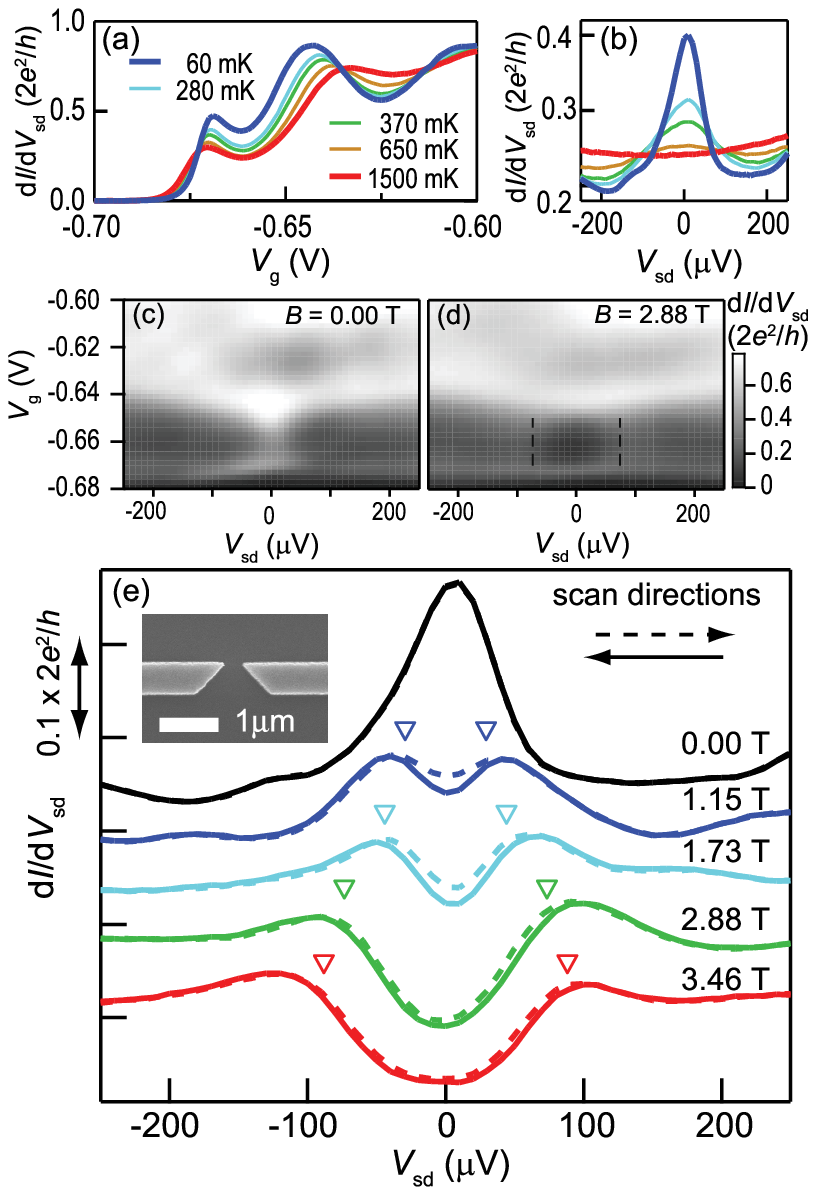}
		\caption{
			\label{fig2}
			(color online).
			(a) Dependence
			of ${\rm d}I/{\rm d}V_{\rm sd}$ on $V_{\rm g}$
			at $V_{\rm sd}$ = 0~$\mu$V
			under temperatures $T$ = 60, 280, 370, 650, and 1500~mK.
			(b) Dependence
			of ${\rm d}I/{\rm d}V_{\rm sd}$ on $V_{\rm sd}$
			at $V_{\rm g}$ =  $-0.66$ V under various temperatures.
			The temperatures are the same as in (a).
			(c) and (d) Gray-scale plot of ${\rm d}I/{\rm d}V_{\rm sd}$
			as a function of $V_{\rm g}$ and $V_{\rm sd}$ for $B$ = 0.00 T (c) and 2.88 T (d).
			Dashed lines in (d) indicate the theoretically expected position
			of ${\rm d}I/{\rm d}V_{\rm sd}$ peaks $\pm |{\rm g}^{*}|\mu_{\rm B}B/e$.
			(e) Dependence of ${\rm d}I/{\rm d}V_{\rm sd}$ on $V_{\rm sd}$ 
			at $V_{\rm g}$ = $-0.66$ V and $T$ = 30~mK under magnetic fields $B$ = 0.00, 1.15, 1.73, 2.88, and 3.46 T.
			The curves are offset by  $0.06 \times 2e^2/h$.
			The expected voltage positions of ${\rm d}I/{\rm d}V_{\rm sd}$ peaks are marked by triangles.
            The solid and dashed curves represent
			the data for the positive  and negative scans, respectively.
			Inset shows scanning electron micrograph of the split-gate device.
			}
	\end{center}
\end{figure}

Figure~2(a)  shows the gate voltage $V_{\rm g}$ dependence of the conductance
of one of the devices. The conductance decreases with decreasing $V_{\rm g}$,
and several conductance peaks appear before the pinch-off,
indicating  formation of a QD at the constriction.
The transport properties of the device 
exhibit behavior typical of the Kondo effect
in a QD\cite{Goldhaber-Gordon, Cronenwett}.
(1)
The conductance shows an increase with decreasing temperature $T$
at a conductance minimum in the ${\rm d}I/{\rm d}V_{\rm sd}$--$V_{\rm g}$ curve
and the opposite temperature dependence 
at the neighboring conductance minimum [Fig.~2(a)].
(2)
A pronounced ZBCP
in the ${\rm d}I/{\rm d}V_{\rm sd}$--$V_{\rm sd}$ curve
develops with decreasing temperature below 700 mK [Fig.~2(b)].
The ZBCPs appear as a bright colored ridge
in the gray-scale plot of ${\rm d}I/{\rm d}V_{\rm sd}$
as a function of $V_{\rm g}$ and $V_{\rm sd}$ [Fig.~2(c)]. 
(3) 
The ${\rm d}I/{\rm d}V_{\rm sd}$ spectrum splits
into two peaks under in-plane magnetic fields $B$ [Fig.~2(e)],
accompanied by the splitting of the ${\rm d}I/{\rm d}V_{\rm sd}$ ridge [Fig.~2(d)].
The peak-to-peak voltage $V_{\rm p-p}$ between the split ${\rm d}I/{\rm d}V_{\rm sd}$ peaks
is almost independent of $V_{\rm g}$ as expected from the theory of the Kondo effect in a QD\cite{Meir}. 
The value of $V_{\rm p-p}$ increases with increasing $B$,
showing an agreement with  theoretically expected values of $V_{\rm p-p} = 2|{\rm g}^{*}|\mu_{\rm B}B/e$
with $|{\rm g}^{*}|$ = 0.44 for GaAs,
as marked by triangles in Fig.~2(e).

\begin{figure}[t]
	\begin{center}
	\includegraphics[width=8.5cm]{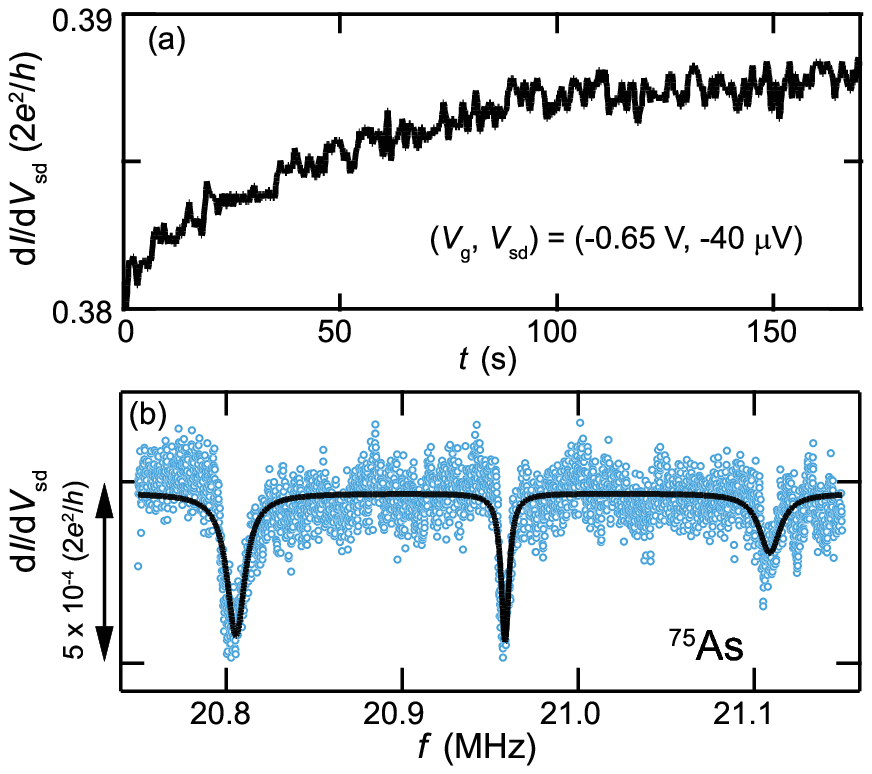}
		\caption{
			(color online).
				(a)
				Temporal change of ${\rm d}I/{\rm d}V_{\rm sd}$ 
				at ($V_{\rm g}$, $V_{\rm sd}$) = ($-$0.65~V, $-$40~$\mu$V) and $B$ = 2.88~T
				after a rapid change in $V_{\rm g}$ and $V_{\rm sd}$ from the pinch-off condition
				($V_{\rm g}$, $V_{\rm sd}$) = ($-$0.70~V, 0~$\mu$V).
				(b)
				NMR spectrum of $^{75}$As obtained by monitoring ${\rm d}I/{\rm d}V_{\rm sd}$
				at ($V_{\rm g}$, $V_{\rm sd}$) = ($-$0.65~V, $-$40~$\mu$V) and $B$ = 2.88~T
				under irradiation of rf-magnetic field.
				Black curve is the Lorentzian fitting result. 
				The splitting of the spectrum
				is due to the electric quadrupole interaction of the nuclei.
		}
	\end{center}
\end{figure}

We find that remarkable hysteresis appears in the ${\rm d}I/{\rm d}V_{\rm sd}$ spectra 
under magnetic fields;
the ${\rm d}I/{\rm d}V_{\rm sd}$--$V_{\rm sd}$ curve of the positive scan
(${\rm d}V_{\rm sd}/{\rm d}t > 0$)  appears above that of the negative scan (${\rm d}V_{\rm sd}/{\rm d}t < 0$)
 as shown in Fig. 2(e).
The ${\rm d}I/{\rm d}V_{\rm sd}$--$V_{\rm sd}$ curves
shown in Fig.~2(e)  are obtained by scanning $V_{\rm sd}$
between $-$500 $\mu$V and $+$500 $\mu$V
in positive and negative directions at a rate of 5 $\mu$V/s.
The hysteresis appears in a particular range of $V_{\rm sd}$
between the ${\rm d}I/{\rm d}V_{\rm sd}$ peaks at each magnetic field.
The hysteresis in the ${\rm d}I/{\rm d}V_{\rm sd}$--$V_{\rm sd}$ curves
indicates involvement of  a phenomenon with a long relaxation time
in the mechanism of electron transport.

Figure 3(a) shows representative data describing
the temporal change in ${\rm d}I/{\rm d}V_{\rm sd}$ at $B$ = 2.88~T
after a rapid change of $V_{\rm g}$ and $V_{\rm sd}$ from the pinch-off condition
to  ($V_{\rm g}$, $V_{\rm sd}$) = ($-$0.65~V, $-$40~$\mu$V) on the slope of the ${\rm d}I/{\rm d}V_{\rm sd}$ peak.
The value of ${\rm d}I/{\rm d}V_{\rm sd}$ increases continuously over a period of 150~s.
The long time required to change ${\rm d}I/{\rm d}V_{\rm sd}$ in Fig. 3(a)
is in accordance with the typical time scale for building up a DNP
in GaAs \cite{Hashimoto, Wald,  Machida, Ren, Kronmuller, Kawamura_APL,  Kawamura_PRB, Ono},
suggesting the occurrence of the DNP in the present QD.

\begin{figure}[t]
	\begin{center}
	\includegraphics[width=8.5cm]{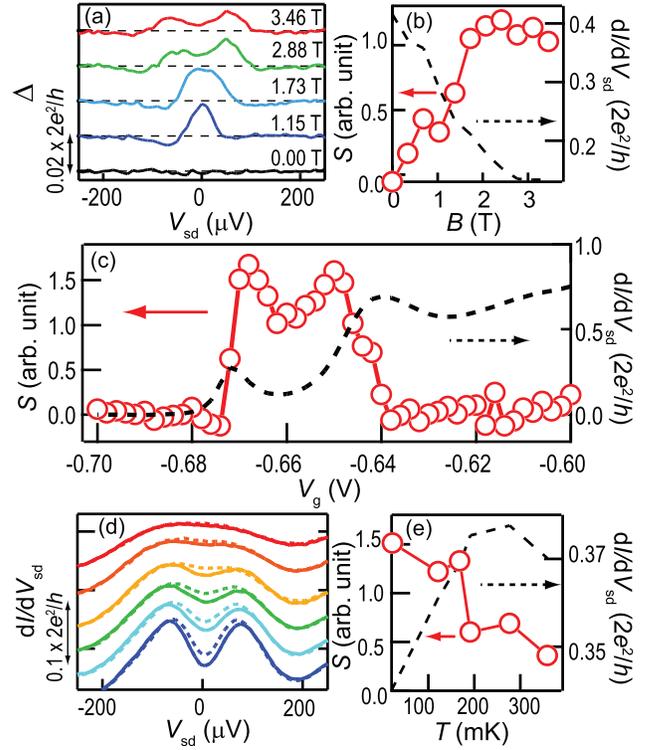}
		\caption{
			\label{fig4}
			(color online).
			(a) Hysteresis $\Delta$ 
			defined as the difference in ${\rm d}I/{\rm d}V_{\rm sd}$ 
			between the positive and negative scans of $V_{\rm sd}$ in Fig.~2(e).
			The  $\Delta$--$V_{\rm sd}$ curves for $V_{\rm g}$ = $-0.66$ V 
			at $B$ = 0.00, 1.15, 1.73, 2.88, and 3.46 T are shown.
			The curves are offset by $0.02 \times 2e^2/h$.
			(b)
			Magnetic field dependences of the area $S$ of the hysteresis (open circles, left axis)
			and ${\rm d}I$/${\rm d}V_{\rm sd} (V_{\rm sd}$ = 0~$\mu$V) (dashed curve, right axis)
			at  $V_{\rm g}$ = $-$0.66~V and $T$ = 30~mK.
			(c)
			Gate voltage dependences of  $S$  (open circles, left axis)
			and ${\rm d}I$/${\rm d}V_{\rm sd} (V_{\rm sd}$ = 0~$\mu$V) (dashed curve, right axis)
			at $B$ = 2.88~T and $T$ = 30~mK.
			(d)
			The ${\rm d}I/{\rm d}V_{\rm sd}$--$V_{\rm sd}$ curves for $V_{\rm g}$ = $-0.65$ V and $B$ = 2.88 T
			under various temperatures $T$ = 60, 120, 170, 190, 280, and 360 mK from bottom to top.
			The solid and dashed curves represent
			the data for the positive  and negative scans, respectively.
			The curves are offset by $0.04 \times 2e^2/h$.
			(e)
			Temperature dependences of  $S$  (open circles, left axis)
			and ${\rm d}I$/${\rm d}V_{\rm sd} (V_{\rm sd}$ = 0~$\mu$V) (dashed curve, right axis)
			at $V_{\rm g}$ = $-$0.65~V and $B$ = 2.88~T.
		}
	\end{center}
\end{figure}

Relevance of the nuclear spins to the hysteresis
and the slow evolution of ${\rm d}I/{\rm d}V_{\rm sd}$
is confirmed by the following NMR spectroscopy measurement.
We apply continuous waves of radio-frequency (rf) magnetic fields 
using a single-turn coil wound around the device.
The frequency $f$ of the rf-magnetic field is scanned 
after the saturation of ${\rm d}I/{\rm d}V_{\rm sd}$
at ($V_{\rm g}$, $V_{\rm sd}$) = ($-$0.65~V, $-$40~$\mu$V).
As shown in Fig.~3(b),
the value of ${\rm d}I/{\rm d}V_{\rm sd}$ decreases when the frequency matches
the NMR frequencies of $^{75}$As
(gyromagnetic ratio $\gamma$ = 45.82~rad$\cdot$MHz/T).
The NMR spectrum of $^{75}$As is obtained\cite{absenceNMR}
by monitoring ${\rm d}I/{\rm d}V_{\rm sd}$.
The NMR spectra of $^{69}$Ga and $^{71}$Ga
are also obtained (not shown).
The observed decreases in ${\rm d}I/{\rm d}V_{\rm sd}$ correspond to 
the decreases in the nuclear spin polarization
due to the absorption of rf-magnetic fields at the NMR frequencies.
Thus, the changes in the nuclear spin polarization are detected
by measuring ${\rm d}I/{\rm d}V_{\rm sd}$
under the Kondo effect regime as expected [Fig. 1(b)].

Because ${\rm d}I/{\rm d}V_{\rm sd}$ reflects changes in the nuclear spin polarization,
the slow increase of  ${\rm d}I/{\rm d}V_{\rm sd}$ in Fig.~3(a) is interpreted as
the development of the DNP in the present QD.
The possibilities of the electron heating or the drift of the gate voltages
are ruled out from the origin of the hysteresis and the slow evolution of  ${\rm d}I/{\rm d}V_{\rm sd}$
because of the absence of the hysteresis at $B$ = 0.00~T.
The induced DNP modifies the ${\rm d}I/{\rm d}V_{\rm sd}$ spectra
through the Zeeman energy,
changing the peak-to-peak voltage to $V_{\rm p-p}  = 2|{\rm g}^{*}|\mu_{\rm B}(B + B_{\rm N})/e$.
Because the DNP develops during the scans of $V_{\rm sd}$
in the ${\rm d}I/{\rm d}V_{\rm sd}$-$V_{\rm sd}$ measurements in Fig.~2(e),
the resultant difference in the nuclear spin polarization at the same $V_{\rm sd}$
causes the hysteresis in the ${\rm d}I/{\rm d}V_{\rm sd}$-$V_{\rm sd}$ curves.
We think that coexistence of the DNP and the Kondo effect
is allowed because
the electron-nuclear flip-flop process of the DNP
does not interfere with the formation of the Kondo singlet significantly
due to the small characteristic energy for the flip-flop process
($\sim$ 1 neV)\cite{Coish, couplingconstant}
compared to that for the Kondo effect ($k_{\rm B}T_{\rm K}$ $\sim$ 80 $\mu$eV).

The above described mechanism of the hysteresis is based
on the sensitivity of the ${\rm d}I/{\rm d}V_{\rm sd}$ spectra
to the changes in the Zeeman energy.
Therefore the conditions for the hysteresis to occur is expected to be related 
to  the appearance of the Kondo effect.
Figure 4(a) shows the hysteresis $\Delta$, defined
as the difference in ${\rm d}I/{\rm d}V_{\rm sd}$
between the positive and negative scans of $V_{\rm sd}$ shown in Fig. 2(e).
To discuss the conditions for the hysteresis to occur,
we introduce the area $S$ of the hysteresis 
by integrating the hysteresis $\Delta$
over a range $-200$~$\mu$V $\le V_{\rm sd} \le$ 200~$\mu$V.
The area $S$, which is zero at $B$ = 0.00~T,
increases with increasing $B$ and becomes saturated above $B$ = 2 T [Fig.~4(b)].
The dependence of $S$ on $V_{\rm g}$ [Fig.~4(c)] reveals that
the hysteresis is remarkable in a range of $V_{\rm g}$
where the Kondo effect appears ($-$0.67~V $< V_{\rm g} <$ $-$0.64~V).
The values of $S$ are almost zero at the other gate voltages.
With increasing temperature $T$ under a fixed magnetic field,
the hysteresis becomes small and the value of $S$ decreases as shown in Figs.~4(d) and 4(e).
The hysteresis almost disappears at around 
$T$ = 360~mK, which is considerably lower than 
the onset temperature of the Kondo effect (about 700~mK).
The temperature where the hysteresis disappears
is close to the temperature 
where the splitting of the ${\rm d}I/{\rm d}V_{\rm sd}$ spectrum becomes unclear [Fig.~4(d)].
These observations
show that the hysteresis is one of the unique features
of the Kondo effect in the magnetic fields
and that the differential conductance 
under the Kondo effect regime is sensitive to the changes
in the effective magnetic field $B + B_{\rm N}$.

\begin{figure}[t]
	\begin{center}
		\includegraphics[width=8.5cm]{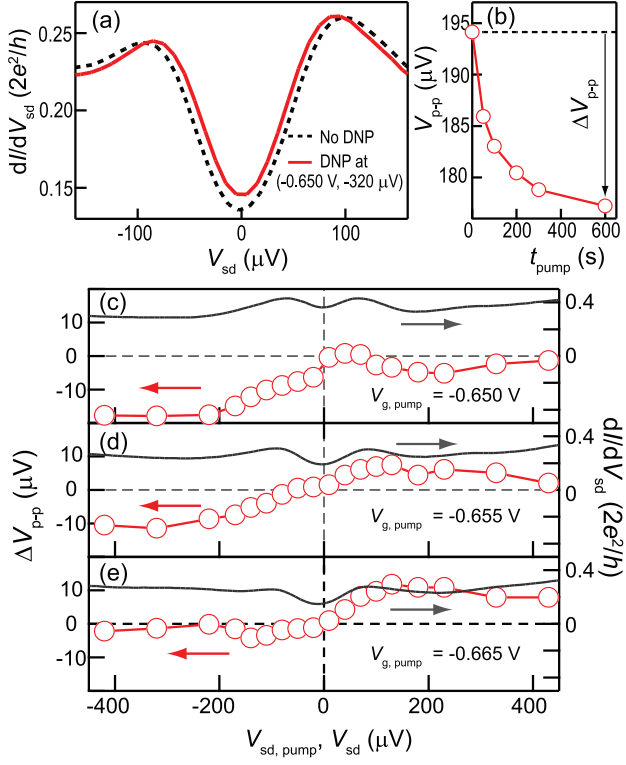}
		\caption{
			(color online).
			(a)
			${\rm d}I/{\rm d}V_{\rm sd}$-$V_{\rm sd}$ curves
			at $V_{\rm g}$ = $-$0.650~V
			obtained without the DNP pumping (dashed)
			and after the DNP pumping (solid) for $t_{\rm pump}$ = 600~s.
			The DNP pumping voltages ($V_{\rm g,pump}$, $V_{\rm sd,pump}$)
			= ($-$0.650~V, $-$320~$\mu$V).
			(b)
			Dependence of $V_{\rm p-p}$ on the DNP pumping period $t_{\rm pump}$.
			The DNP pumping voltages are the same as in (a).
			The DNP is relaxed between each measurement
			by waiting more than 20 minutes under the pinch-off condition.
			(c)-(e)
			Dependence of $\Delta V_{\rm p-p}$
			on the DNP pumping voltages ($V_{\rm g,pump}$, $V_{\rm sd,pump}$)
			for $t_{\rm pump}$ = 600 s.
			The $V_{\rm sd}$ dependences of ${\rm d}I/{\rm d}V_{\rm sd}$ (solid curve)
			are plotted on the right axis.
			}
	\end{center}
\end{figure}

The effect of the DNP on the ${\rm d}I/{\rm d}V_{\rm sd}$ spectra
can be seen more directly in the following pump-probe measurement. 
The dashed and solid curves in Fig.~5(a) show the 
${\rm d}I/{\rm d}V_{\rm sd}$--$V_{\rm sd}$ curves
at the same $V_{\rm g}$.
The dashed curve is obtained immediately
after relaxing the nuclear spin polarization under the pinch-off condition
and the solid curve is obtained after the DNP pumping
at ($V_{\rm g,pump}$, $V_{\rm sd,pump}$) = ($-$0.650~V, $-$320~$\mu$V)
for $t_{\rm pump}$ = 600 s.
The peak-to-peak voltage $V_{\rm p-p}$ after the DNP pumping is smaller 
than that without the DNP pumping.
With increasing the period of the DNP pumping $t_{\rm pump}$, 
 $V_{\rm p-p}$ decreases gradually, as shown in Fig.~5(b).
The gradual change in $V_{\rm p-p}$ with increasing $t_{\rm pump}$
gives an additional evidence for the occurrence of the DNP.
The shift $\Delta V_{\rm p-p}$
measured from the value of $V_{\rm p-p}$ without the DNP pumping
allows us to evaluate the effective magnetic field $B_{\rm N}$
using a proportional relation $\Delta V_{\rm p-p}/V_{\rm p-p} = B_{\rm N}/(B+B_{\rm N})$,
which is yielded based on an assumption $V_{\rm p-p}  = 2|{\rm g}^{*}|\mu_{\rm B}(B+B_{\rm N})/e$.
A representative value of $\Delta V_{\rm p-p}$ = $-$17.9~$\mu$V 
obtained from the difference between the dashed and solid curves in Fig.~5(a)
corresponds to $B_{\rm N}$ = $-$0.26~T.
This value is equivalent to the nuclear spin polarization of 4.9~\%
using the expression for $B_{\rm N}$ in GaAs obtained in Ref. \onlinecite{Paget}.

Figures~5(c)-5(e) show the dependence of $\Delta V_{\rm p-p}$
on the DNP pumping voltages ($V_{\rm g,pump}$, $V_{\rm sd,pump}$).
The values of $\Delta V_{\rm p-p}$, which are proportional to $B_{\rm N}$,
are almost zero at around $V_{\rm sd, pump}$ = 0, 
while large values of $|\Delta V_{\rm p-p}|$ are obtained at $|V_{\rm sd}| > 100$ $\mu$V.
In addition, the sign of $\Delta V_{\rm p-p}$,
which indicates the DNP polarity, strongly depends on the DNP pumping voltage.
The negative $\Delta V_{\rm p-p}$ ($B_{\rm N} < 0$) is remarkable at $V_{\rm g, pump}$ = $-0.650$ V
with $V_{\rm sd, pump} < 0 $ [Fig. 5(c)], while the positive $\Delta V_{\rm p-p}$ ($B_{\rm N} > 0$)
is remarkable at  $V_{\rm g, pump}$ = $-0.665$ V with $V_{\rm sd, pump} > 0$ [Fig. 5(e)].
At the gate voltage between them ($V_{\rm g, pump}$ = $-0.655$ V),
both polarities appear depending on the sign of $V_{\rm sd, pump}$ [Fig. 5(d)].

The optimum values of $|V_{\rm sd, pump}|$ for the DNP pumping
are larger than the $|V_{\rm sd}|$ values for the ${\rm d}I/{\rm d}V_{\rm sd}$ peaks.
Moreover, the $V_{\rm sd, pump}$ range where the DNP is achieved
does not coincide with the $V_{\rm sd}$ range where the hysteresis is observed.
Both facts suggest that the mechanism for the DNP pumping
is not directly related to that of the Kondo effect.
In addition, the  dependence of the DNP polarity 
on the sign of $V_{\rm sd, pump}$ indicates that 
the present QD is asymmetric
and that the asymmetry plays an important role in the DNP pumping mechanism.
We think that the positive $B_{\rm N}$ could be attributed to the electron spin relaxation in the QD.
When adding an electron to the empty QD, both up-spin and down-spin state are allowed.
If the added electron is in the down-spin state,
it can relax to the up-spin state in the QD before going to the electrode.
The down-to-up flips of  electron spins
cause to decrease the nuclear spin  polarization $\langle I_z \rangle$
via the electron-nuclear flip-flop process
of the hyperfine interaction, building up the positive $B_{\rm N}$
in Fig. 5(e)\cite{direction}.
Because of the asymmetric couplings between the QD and the electrodes,
 the added electron probably goes through  the QD
before relaxing its spin when  $V_{\rm sd, pump} < 0$.
The negative $B_{\rm N}$ in Fig. 5(c) indicates involvement of 
the up-to-down flips of electron spins in the QD.
At the present stage, we do not have a comprehensive explanation
why the up-to-down flips become dominant as the gate voltage is increased.
Further studies will be needed to fully understand the DNP pumping mechanism.

\begin{figure}[t]
	\begin{center}
		\includegraphics[width=8.5cm]{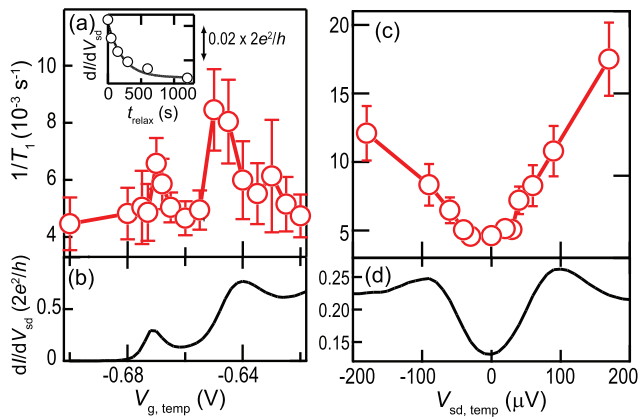}
		\caption{
			(color online).
			(a) Dependence of $1/T_{1}$ on $V_{\rm g, temp}$ for $V_{\rm sd, temp}$ = 0~$\mu$V and $B$ = 2.88~T.
			Inset shows a representative decay curve of the nuclear spin polarization
			at the pinch-off condition ($V_{\rm g, temp}$, $V_{\rm sd, temp}$) = ($-0.70$ V, 0 $\mu$V).
			(b) ${\rm d}I/{\rm d}V_{\rm sd}$-$V_{\rm g}$ curve at $V_{\rm sd}$ = 0 $\mu$V
			(c)  Dependence of $1/T_{1}$ on $V_{\rm sd, temp}$ 
			for $V_{\rm g, temp}$ = $-0.66$~V and $B$ = 2.88~T.
			(d) ${\rm d}I/{\rm d}V_{\rm sd}$-$V_{\rm sd}$ curve at $V_{\rm g}$ = $-0.66$~V.
			}
	\end{center}
\end{figure}

Impact of  the newly developed techniques
of the DNP and the resistive detection of nuclear spins
is further emphasized by the following
nuclear spin relaxation rate  $1/T_1$ measurements.
Because $1/T_1$ is enhanced by the electron spin fluctuation,
the electron spin dynamics in the QD can be studied through the $1/T_1$ measurements.
The decay curve of nuclear spin polarization is obtained by the following procedure:
First, the DNP is pumped at ($V_{\rm g}$, $V_{\rm sd}$) = ($-0.650$ V, $-320$ $\mu$V).
Then, the nuclear spin polarization is relaxed under a certain voltage condition ($V_{\rm g, temp}$, $V_{\rm sd, temp}$)
for a given time $t_{\rm relax}$.
Finally, the change in the nuclear spin polarization is read out by monitoring ${\rm d}I/{\rm d}V_{\rm sd}$
at ($-0.650$ V, 0 $\mu$V).
By repeating this procedure with different periods $t_{\rm relax}$, we obtain the decay curve
of the nuclear spin polarization as shown in the inset of Fig. 6(a).
The value of $1/T_1$ is evaluated by fitting the decay curve to the form
${\rm d}I/{\rm d}V_{\rm sd}  = a + b\exp (-t_{\rm relax}/T_1)$.

Figure 6(a) shows the dependence of $1/T_{1}$
on $V_{\rm g, temp}$ at $B$ = 2.88~T for $V_{\rm sd, temp}$ = 0~$\mu$V.
The relaxation rate $1/T_{1}$ = 4.4 $\times$ 10$^{-3}$~s$^{-1}$
for $V_{\rm g, temp}$ = $-$0.70~V
(the QD is depleted during the relaxation) 
is in the same order as the reported  $1/T_{1}$ values in GaAs\cite{Hashimoto}.
The relaxation rate increases as $V_{\rm g, temp}$ approaches
the conductance peaks [Fig. 6(b)].
We think that electron spin fluctuations introduced 
by the electron tunneling cause the increase in $1/T_{1}$.
The small values  of  $1/T_{1}$ at around $V_{\rm g, temp}$ = $-0.66$~V 
can be attributed to the suppression of the electron spin fluctuation due to the Zeeman effect.
When a bias voltage $V_{\rm sd, temp}$ is applied during the relaxation [Fig. 6(c)],
the values of  $1/T_{1}$ do not increase largely at $|V_{\rm sd, temp}| <$ 40 $\mu$V.
$1/T_{1}$ becomes large at  $|V_{\rm sd, temp}| >$ 40 $\mu$V
associated with the increase in ${\rm d}I/{\rm d}V_{\rm sd}$ [Fig. 6(d)].
The increase in $1/T_{1}$ is probably related to
the electron transport through the Kondo  resonance state.
We think that the $1/T_{1}$ measurement using the DNP technique
is useful to understand spin properties of the Kondo effect under the non-equilibrium regime.

This work was supported by the PRESTO of JST,
Grant-in-Aid for Scientific Research from the MEXT,
and the Project for Developing Innovation Systems of the MEXT.
We thank M. Eto, T. Kubo, S. Amaha, and E. Minamitani for helpful discussions.
D. G. acknowledges the RIKEN-Konstantz Univ. internship program.


\begin{thebibliography}{99}
\bibitem{absenceNMR}
	NMR spectra are not obtained at bias and gate voltage conditions
	where the hysteresis is not observed.
\bibitem{couplingconstant}
	We assume the averaged hyperfine coupling constant 
	$A$ = 90 $\mu$eV for GaAs\cite{Paget}
	and 10$^5$ nuclei in the QD.
\bibitem{direction}
	Because  ${\rm g}^* < 0$ and $A >0 $ in GaAs,
	a negative nuclear spin polarization $\langle I_z \rangle < 0$ corresponds to  $B_{\rm N} > 0$.
\end{thebibliography}
\end{document}